**R.Decarli[1], R.Falomo[2], J.Kotilainen[3], M.Labita[1], R.Scarpa[4], A.Treves[1]**

[1]Università degli Studi dell'Insubria, via Valleggio 11, I-22100 Como, Italy
[2]Osservatorio Astronomico di Padova, vicolo dell'Osservatorio 5, I-35122, Padova, Italy
[3]Tuorla Observatory, Department of Physics and Astronomy, Unviersity of Turku, Väisäläntie 20, FI-21500 Piikkiö, Finland
[4]Instituto de Astrofisica de Canarias, via Lactea s/n, E-38205, La Laguna, Spain

# Re-classification of the alleged quasar Q0045-3337

## Abstract

We present a medium-resolution optical spectrum of the alleged high-redshift quasar Q0045-3337, taken at the ESO/3.6m telescope. Our observations show that the object is not a quasar but a star of spectral type B. We suggest that the object is either a white dwarf or a halo population Blue Horizontal Branch star.

## Introduction

We report on optical spectroscopic observations of the source Q0045-3337 ($RA_{J2000}$: 00 47 41.8, $DEC_{J2000}$: -33 20 55), that was classified [1] as a quasar (object 004517.0-333717). In that work, the Automated Quasar Detection technique [2] was applied to objective prism plates. Due to the poor spectral resolution, the redshift (z=2.14) was determined only tentatively. The object is not flagged as a ``good quality candidate'', and no X-ray or radio counterparts were found. The target appears also in the Veron-Cetty & Veron quasar catalog [3] (J004741.9-332055; V=18.75).

The source was imaged [4] in the Ks band with VLT/NACO using the adaptive optics technique, in a campaign aimed to observe host galaxies of high-z quasars that lie in the vicinity of bright stars [4],[5]. No host galaxy was detected around the target, which appeared with Ks=18.75 mag. A foreground spiral galaxy of unknown redshift and Ks=17.5 mag was observed at ~1.2 arcsec from the target. Since the separation between Q0045-3337 and the galaxy is comparable to the estimated Einstein radius of the galaxy, it was argued that Q0045-3337 could be gravitationally lensed [4],[6]. Stimulated by these considerations we obtained medium-resolution spectroscopy of the source in order to verify the reported redshift of Q0045-3337.

## Observations and analysis

The object was observed at the ESO/3.6m telescope on 2007 September, 12, as a part of a spectroscopic study of high-redshift quasars (ESO proposal ID: 079.B-0304(A)). The ESO Faint Object Spectrograph and Camera [7] was mounted in long-slit spectroscopy setup with a grism yielding $\lambda/\Delta\lambda$ ~400 and a 1.2 arcsec slit. The observed wavelength range is 4100-7500 Ang. The wavelength calibration accuracy is ~0.3 Ang throughout the whole range. At $\lambda$~5000 Ang, the spectral resolution is ~13 Ang. A one-dimensional spectrum was integrated over a ~2 arcsec aperture. The signal-to-noise ratio at 5500 Ang is ~60.

We also obtained Johnson's V- and R-band images of the source. Figure 1 shows the R-band field of the target. The astrometric solution was computed through the positions of GSC2 catalog sources. The target coordinates in the NASA/IPAC Extragalactic Database are marked, leaving no doubt concerning the identification of the source.

Three 800 s exposures were taken. The standard IRAF procedure was adopted in the data reduction. The *ccdred* package was employed to perform bias subtraction, flat field correction, image alignment and combination. The spectrum extraction, background subtraction and the calibrations both in wavelength and flux were performed with the *doslit* task in the *kpnoslit* package, using a He-Ar lamp and a standard spectrophotometric star as a reference. Absolute flux calibration of spectra was corrected through the photometry of field stars, by comparing corollary imaging with R filter to the magnitudes published in the GSC2 catalog. The measured R-band magnitude of the target (blended to the nearby galaxy) is ~18.5 ± 0.1. Galactic extinction was accounted for [8] (E(B-V)=0.017 mag), assuming $R_V$=3.1.

The slit was oriented along the direction between the target and the companion galaxy. The seeing was 1.4 arcsec and therefore the spectrum is the blend of these sources. On the other hand, the R-band magnitude expected for the close-by galaxy is 20.1 (assuming R-K=2.6 mag [9]), that is its contribution to the observed flux is negligible. Neither the V- and R-band images nor the spatial profile of the spectrum show any signature of the galaxy, supporting the idea that the blended spectrum is dominated by Q0045-3337.

## *Results and conclusions*

The optical spectrum, presented in figure 2, clearly shows the presence of rest-frame hydrogen absorption features over a blue continuum, indicating that it is a star of spectral type B [10]. We estimate a radial velocity of 74 ± 16 km/s, corrected to the Galactic Center ($b$=-83.7 degrees). Given the observed magnitude $m_V$=18.7, if we assume $M_V$=-3, typical of a main sequence B star, the distance of the Q0045-3337 would be 220 kpc, excluding that it is a main sequence star of the Galaxy. We consider two other possibilities: 1) that the source is a white dwarf with 8000 K and log $g$ =7.5 - 8.0; 2) a Blue Horizontal Branch (BHB) star of the halo. In the former case, assuming an absolute magnitude of $M_V$ = 12-13, the inferred distance would be about 150 pc. In the latter case, assuming $M_V$=+0.5, the distance would be ~45 kpc. It is noticeable that, according to the Sloan Digital Sky Survey [11], the surface density of quasars candidates at g~18.7 (~4 per magnitude in a square degree[12]) is ~4 times that of white dwarfs [13], and ~40 times the surface density of BHB candidates [14].

In conclusion, we have shown that Q0045-3337 is not a quasar, but likely a white dwarf. Other similar cases were discussed in [15] and [16].

## *Acknowledgments*

We thank Angela Iovino and Massimo Dotti for useful discussions.

## *References*

[1] Iovino A., Clowes R.G., Shaver P., A large sample of objective prism quasar candidates, Astron. & Astroph. Suppl., 1996; 119: 265
[2] Clowes R.G., Automated quasar detection in the SGP field – A clustering study, Mon. Not. Royal Astron. Soc., 1986: 218: 139
[3] Veron-Cetty M.-P., Veron P., A catalogue of quasars and active nuclei: 10$^{th}$ edition, Astron. & Astrop., 2001; 374: 92
[4] Falomo R., Kotilainen J.K., Scarpa R., Treves A., VLT adaptive optics imaging of QSO host galaxies and their close environment at z~2.5: Results from a pilot program, Astron. & Astrop., 2005; 434: 469
[5] Falomo R., Treves A., Kotilainen J.K., Scarpa R., Uslenghi M., Near-Infrared Adaptive Optics Imaging of High-Redshift Quasars, Astroph. Journ., 2008; 673: 694
[6] Chieregato M., Miranda M., Jetzer P., Q0045-3337: models including strong lensing by a spiral galaxy, Astron. & Astroph., 2007; 474: 777
[7] Buzzoni B., Delabre B., Dekker H., *et al.*, The ESO Faint Object Spectrograph and Camera (EFOSC), ESO Messenger 1984; 38: 9
[8] Schlegel *et al.*, Maps of dust infrared emission for use in estimation of reddening and Cosmic Microwave Background radiation foregrounds, Astroph. Journ., 1998; 500: 525
[9] Buzzoni A., Broad-band colours and overall photometric properties of template galaxy models from stellar population synthesis, Mon. Not. Royal Astron. Soc., 2005; 361: 725
[10] Allen C.W., 1973, Astrophysical Quantities, Athlone (3$^{rd}$ edition), London 1976
[11] Adelman-McCarthy J.K., Agüeros M.A., Allam S.S., Allende Prieto C., *et al.*, The sixth Data Release of the Sloan Digital Sky Survey, Astroph. Journ. Suppl., 2008; 175: 297
[12] Yanny B., Newberg H.J., Kent S., Laurent-Muehleisen S.A., Pier J.R., *et al.*, Identification of A-colored stars and structure in the halo of the Milky Way from Sloan Digital Sky Survey Commissioning Data, Astroph. Journ., 2000; 540: 825
[13] Bianchi L., Rodriguez-Merino L., Viton M., Laget M., Efremova B., *et al.*, Statistical properties of the GALEX-SDSS matched source catalogs, and classification of the UV sources, Astroph. Journ. Suppl., 2007; 173: 659
[14] Pier J.R., Finding BHB stars with SDSS to contrain HVC distances, Astron. Society Pacific Conf., 1999; 166: 73
[15] Sbarufatti B., Treves A., Falomo R., Heidt J., Kotilainen J., Scarpa, R., ESO Very Large Telescope Optical Spectroscopy of BL Lacertae Objects. II New redshifts, featureless objects and classification assessments, Astron. Journ.,

2006; 132: 1

[16] De Martino D., Koester D., Treves A., Sbarufatti B., Falomo R., An extremely Carbon-rich White Dwarf in the direction of the Virgo-Coma cluster, Astron. Society Pacific Conf., 2007; 372: 273

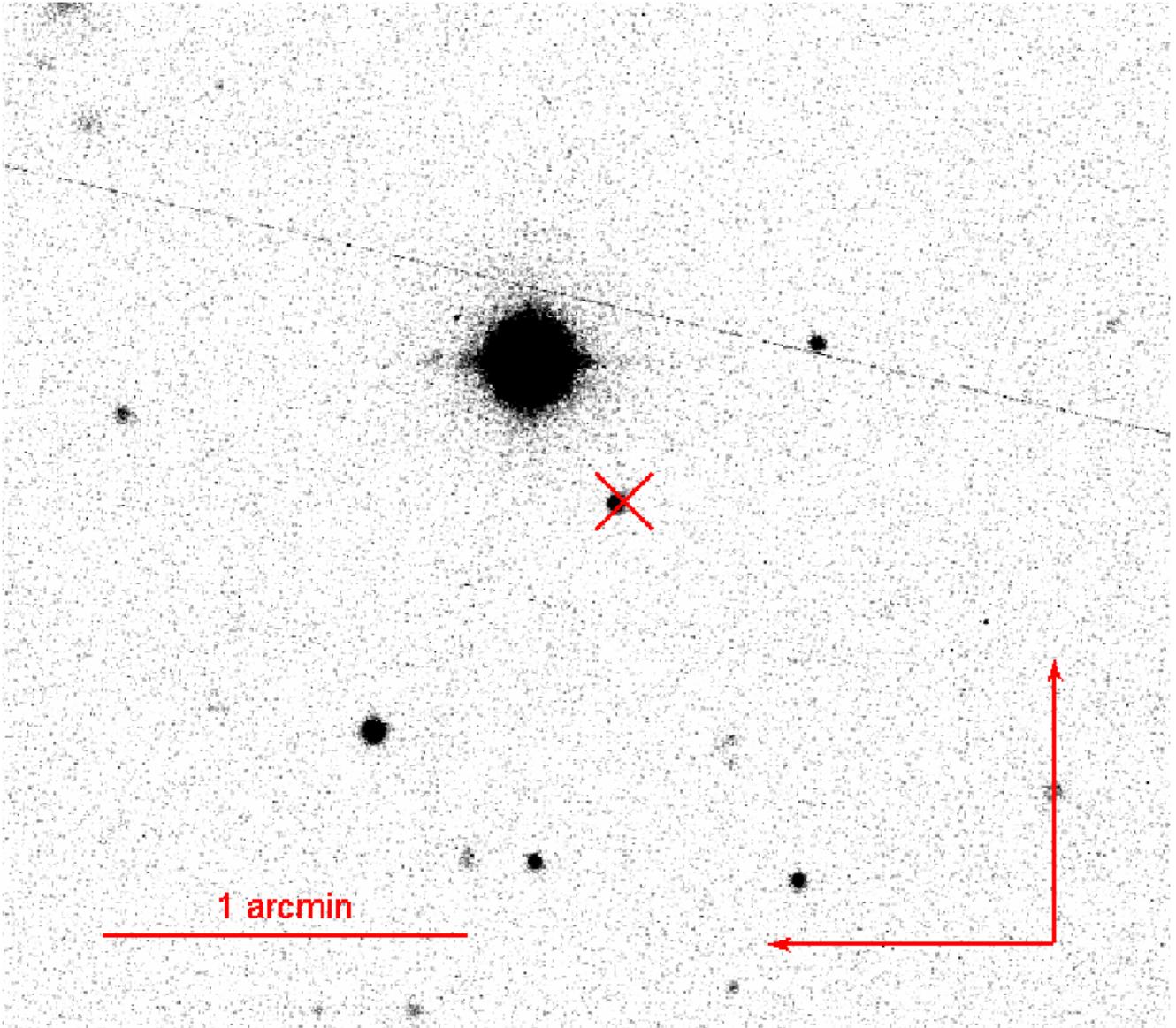

Figure 1 - The field of Q0045-3337, as imaged with the R-band filter. North is up, East is to the left-hand side of the frame. The `x' symbol marks the corresponding coordinates in the NASA/IPAC Extragalactic Database.

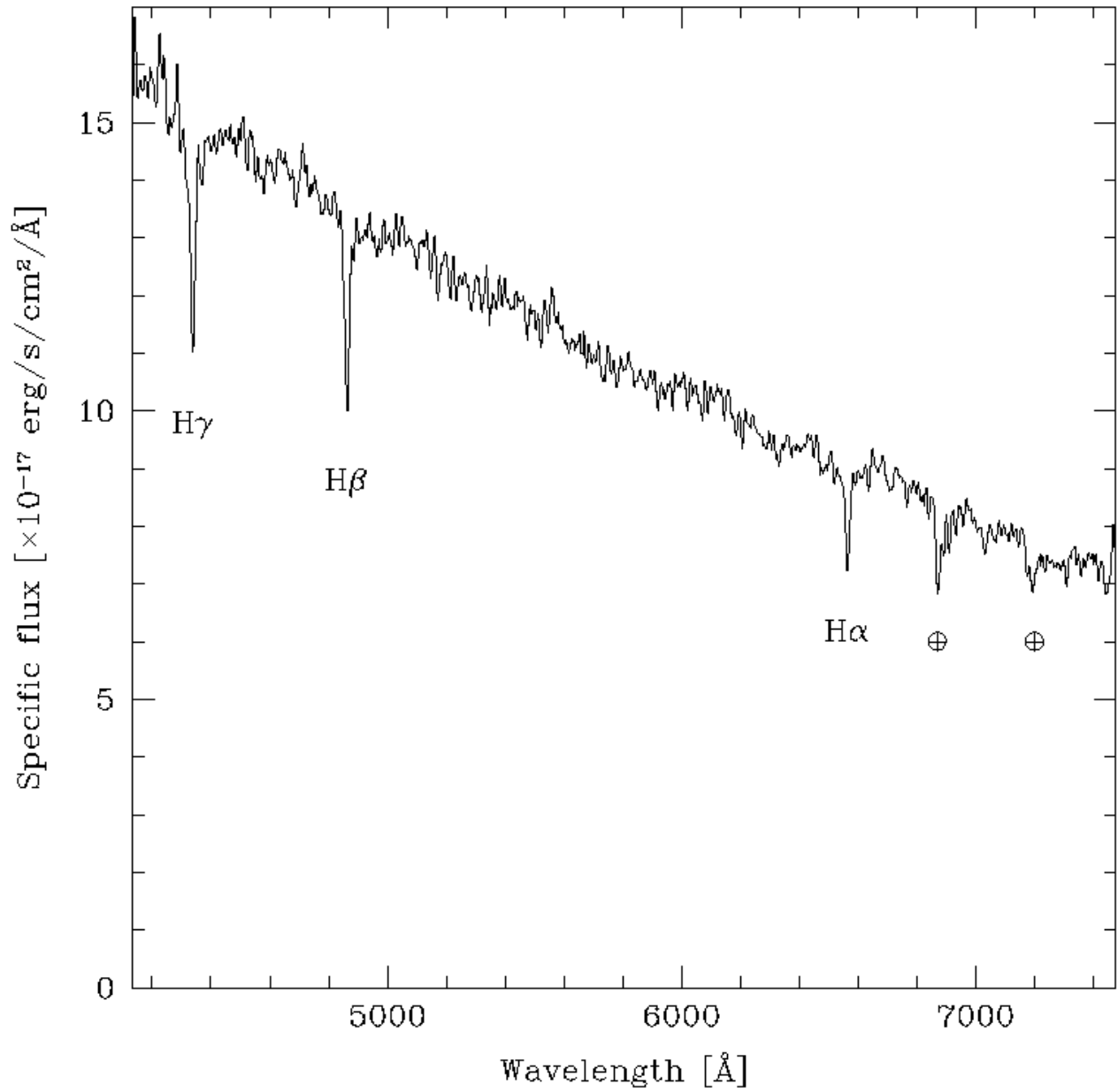

The optical spectrum of the alleged quasar Q0045-3337. The rest-frame Balmer absorption lines indicate that the object is a B-type star. The Earth symbol marks the main atmospheric absorption lines.